\documentclass[aps,showpacs,superscriptaddress,twocolumn]{revtex4}%
\usepackage{amsfonts}
\usepackage{amsmath}
\usepackage{amssymb}
\usepackage{graphicx}%
\begin{document}
\title{Topological edge states and quantum Hall effect in the Haldane model}
\author{Ningning Hao}
\affiliation{Institute of Physics, Chinese Academy of Sciences, Beijing 100080, People's
Republic of China}
\author{Ping Zhang}
\thanks{E-mail: zhang\_ping@iapcm.ac.cn}
\affiliation{Institute of Applied Physics and Computational
Mathematics, P.O. Box 8009, Beijing 100088, People's Republic of
China} \affiliation{Center for Applied Physics and Technology,
Peking University, Beijing 100871, People's Republic of China}
\author{Zhigang Wang}
\affiliation{Institute of Applied Physics and Computational
Mathematics, P.O. Box 8009, Beijing 100088, People's Republic of
China}
\author{Wei Zhang}
\affiliation{Institute of Applied Physics and Computational
Mathematics, P.O. Box 8009, Beijing 100088, People's Republic of
China}
\author{Yupeng Wang}
\affiliation{Institute of Physics, Chinese Academy of Sciences, Beijing 100080, People's
Republic of China}

\begin{abstract}
We study the topological edge states of the Haldane's graphene model with
zigzag/armchair lattice edges. The Harper equation for solving the energies of
the edge states is derived. The results show that there are two edge states in
the bulk energy gap, corresponding to the two zero points of the Bloch
function on the complex-energy Riemann surface. The edge-state energy loops
move around the hole of the Riemann surface in appropriate system parameter
regimes. The quantized Hall conductance can be expressed by the winding
numbers of the edge states, which reflects the topological feature of the
Haldane model.

\end{abstract}
\maketitle

The integer quantum Hall effect (IQHE), discovered in 1980 by Klaus von
Klitzing \cite{Klitzing}, is a striking set of macroscopic quantum phenomena
observed in a high mobility two-dimensional electron gas (2DEG) in a strong
transverse magnetic field (typically, $B\sim1$--$30$T). Soon after the
Laughlin's famous gauge invariance argument and the treatment of the IQHE as
an adiabatic quantum pump \cite{Laughlin}, it was shortly recognized
\cite{Tho1982} that the Hall conductance $\sigma_{xy}$ at the plateaus can be
understood in terms of topological invariants known as Chern numbers
\cite{Chern}, which are integrals of the $k$-space Berry curvatures of the
bulk states over the magnetic Brillouin zone. While IQHE finds its elegant
connection through the adiabatic curvature with bulk topological invariants,
Halperin \cite{Halperin} first stressed that the existence of the sample
edges, which produces the current-carrying localized edge states in the Landau
energy gap, is essential in the Laughlin's gauge invariance argument. Hatsugai
further developed a topological theory of the edge states \cite{Hatsugai}, in
which topological invariants are the winding numbers of the edge states on the
complex-energy Riemann surface (RS).

The presence of IQHE fundamentally rely on the breaking of the time-reversal
symmetry (TRS), which in the above mentioned works is brought about by
imposing an external magnetic field on the electrons. Besides this external
magnetic field, the TRS also can be broken by a variety of the other extrinsic
or intrinsic mechanisms. A most straightforward way is, like what has been
carried out in the Aharonov-Bohm (AB) effect, the introduction of magnetic
flux (instead of magnetic field) to the Bloch electrons. In virtue of such a
way, by using a graphene lattice with the complex hopping matrix elements of
the next-nearest-neighboring honeycomb sites included, Haldane first showed
that the non-zero Chern numbers and thus the IQHE can be realized in zero
magnetic field \cite{Haldane}. In contrast with the cases with external
magnetic field, a detailed study of the topological edge states of the Haldane
model is still lacking. This issue is stressed in the present paper. Also, our
study is motivated by the observation that in addition to its importance in
charge IQHE, the spin-doubled Haldane model, in which TRS is recovered, has
also played a key role in understanding the quantum spin Hall effect (QSHE)
\cite{Kane1} and new phase of matter \cite{Kane2}.

Our discussion of the topological edge states for the Haldane model begins
with deriving the Harper equation to describe the wave-function transfer
relation between two edges in a graphene ribbon. It is found that there are
two edge states in the bulk energy gap, corresponding to two zero points of
the Bloch function on the complex-energy RS. The edge-state energy loops move
around the hole of the RS, giving rise to nontrivial winding numbers. The
quantized Hall conductance can be expressed by the winding numbers of the edge
states, which reflects the topological feature of the Haldane model.

The graphene lattice is composed of two sublattices (denoted by the
red and blue dots in Fig. 1). The graphene ribbons with zigzag edges
and armchair edges are plotted in Fig. 1(a) and 1(b), respectively.
The lattice tight-binding Hamiltonian \cite{Haldane} is given by
\begin{equation}
H=\underset{i}{\sum}t_{0}c_{i}^{\dag}c_{i}+\underset{\langle i,j\rangle}{\sum
}t_{1}c_{i}^{\dag}c_{j}+\underset{\langle\langle i,j\rangle\rangle}{\sum}%
t_{2}e^{i\phi_{ij}}c_{i}^{\dag}c_{j}. \tag{1}\label{e1}%
\end{equation}
In the above Hamiltonian, the on-site energy $t_{0}$=$+M$\ on A\ site and
$-M$\ on B site. $t_{1}$ and $t_{2}$\ are real hopping matrix elements between
nearest neighbors on the different and the same sublattices, respectively. To
break TRS, a complex phase $\phi_{ij}$ is introduced to the next nearest
neighbor hopping $t_{2}$. Following Haldane \cite{Haldane}, we set the
magnitude of this complex phase as $|\phi_{ij}|=\phi$, and the direction of
the positive phase is anticlockwise. Note that the net flux is zero in one
unit cell. Since the spin-orbital effect is not included, we neglect the spin
indices for simplicity.

\begin{figure}[ptb]
\begin{center}
\includegraphics[width=1.0\linewidth]{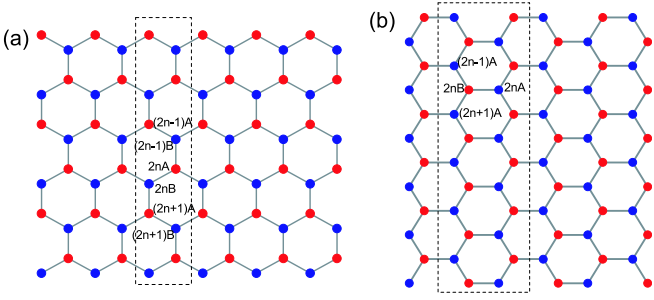}
\end{center}
\caption{(Color online) The structure of graphene ribbons with (a) zigzag
edges and (b) armchair edges. The rectangle with the dashed line is the unit
cell.}%
\end{figure}

Now, let us derive the Harper equation \cite{Harper, Hosfstadter} of the
graphene ribbons with zigzag edges. We suppose that the system is periodic in
the $x$ direction while it has two edges in the $y$\ direction [see Fig.
1(a)]. In the following we replace index $i$\ with $\left(  mns\right)  $\ to
denote the lattice site, where $\left(  mn\right)  $\ label the unit cells and
$s$\ label the sites A and B in this cell. The distance of the nearest
neighboring lattice sites is set to be unity throughout this paper. Since the
lattice is periodic in the $x$ direction, we can use a momentum representation
of the electron operator%

\begin{equation}
c_{mns}=\frac{1}{\sqrt{L_{x}}}\underset{k_{x}}{\sum}e^{ik_{x}X_{mns}}%
c_{ns}\left(  k_{x}\right)  , \tag{2}\label{e2}%
\end{equation}
where $\left(  X_{mns},Y_{mns}\right)  $\ represents the coordinate of the
site $s$\ in the unit cell $\left(  mn\right)  $, and $k_{x}$ is the momentum
along the $x$ direction. Let us consider the one-particle state $\left\vert
\mathbf{\psi}(k_{x})\right\rangle $=$\sum\mathbf{\psi}_{ns}(k_{x})c_{ns}%
^{\dag}(k_{x})\left\vert 0\right\rangle $. Inserting it into the
Schr\"{o}dinger equation $H\left\vert \mathbf{\psi}\right\rangle $%
=$\epsilon\left\vert \mathbf{\psi}\right\rangle $, one can easily get the
following two eigenvalue equations for sites A and B:%
\begin{align}
\epsilon\mathbf{\psi}_{nA}  &  =\left(  M+f_{+}\right)  \mathbf{\psi}%
_{nA}+g_{-}(\mathbf{\psi}_{(n+1)A}+\mathbf{\psi}_{(n-1)A})\nonumber\\
&  +t_{1}\mathbf{\psi}_{(n-1)B}+g_{0}\mathbf{\psi}_{nB},\nonumber\\
\epsilon\mathbf{\psi}_{nB}  &  =(-M+f_{-})\mathbf{\psi}_{nB}+g_{+}%
(\mathbf{\psi}_{(n+1)B}+\mathbf{\psi}_{(n-1)B})\nonumber\\
&  +t_{1}\mathbf{\psi}_{(n+1)A}+g_{0}\mathbf{\psi}_{nA}, \tag{3}\label{e3}%
\end{align}
where $f_{\pm}$=$2t_{2}\cos\left(  \sqrt{3}k_{x}\pm\phi\right)  $, $g_{\pm}%
$=$2t_{2}\cos\left(  \frac{\sqrt{3}}{2}k_{x}\pm\phi\right)  $, and $g_{0}%
$=$2t_{1}\cos\left(  \frac{\sqrt{3}}{2}k_{x}\right)  $. Eliminating the B
sites, we obtain a difference equation for A sites%

\begin{equation}
\mathbf{\psi}_{n}=b(\mathbf{\psi}_{(n-1)}+\mathbf{\psi}_{(n-3)})+d\mathbf{\psi
}_{(n-2)}-\mathbf{\psi}_{(n-4)}, \tag{4}\label{e4}%
\end{equation}
where%
\begin{align}
b  &  =\frac{1}{g_{+}g_{-}}\left\{  2(g_{+}+g_{-})\epsilon+2g_{0}\left[
t_{1}-\frac{t_{2}^{2}}{t_{1}}\right]  \right. \nonumber\\
&  \left.  +\frac{t_{2}}{t_{1}}g_{0}M\sin\phi-4t_{2}^{2}\cos\left(
\frac{3\sqrt{3}}{2}k_{x}\right)  \cos2\phi\right\}  ,\nonumber\\
d  &  =\frac{1}{g_{+}g_{-}}\left\{  2\left(  M^{2}+t_{1}^{2}-\epsilon
^{2}\right)  -f_{+}f_{-}\right. \tag{5}\label{e5}\\
&  \left.  +4t_{2}\left[  \frac{g_{0}^{2}}{t_{1}^{2}}-2\right]  \left(
\epsilon\cos\phi-M\sin\phi\right)  \right\}  -2,\nonumber
\end{align}
and $\mathbf{\psi}_{nA}$ was replaced by $\mathbf{\psi}_{n}$. Eq. (\ref{e4})
is the so-called Harper equation \cite{Harper, Hosfstadter}. The next key step
is to represent Eq. (\ref{e4}) in the transfer matrix form. After a tedious
but straightforward derivation, we find that by introduction of a new wave
function $\mathbf{\varphi}_{n}$, which is a linear transformation of the
original wave function $\mathbf{\psi}_{n}$,%

\begin{equation}
\mathbf{\varphi}_{n}=\mathbf{\psi}_{n}+\frac{-b+\sqrt{b^{2}+4(2+d)}}%
{2}\mathbf{\psi}_{n-1}+\mathbf{\psi}_{n-2}, \tag{6}\label{e7}%
\end{equation}
then the new wave function $\mathbf{\varphi}_{n}$\ can be written in the
following transfer matrix form%
\begin{equation}
\left(
\begin{array}
[c]{c}%
\mathbf{\varphi}_{n}\\
\mathbf{\varphi}_{n-1}%
\end{array}
\right)  =\left(
\begin{array}
[c]{cc}%
t & -1\\
1 & 0
\end{array}
\right)  \left(
\begin{array}
[c]{c}%
\mathbf{\varphi}_{n-1}\\
\mathbf{\varphi}_{n-2}%
\end{array}
\right)  \equiv\widetilde{M}(\epsilon)\left(
\begin{array}
[c]{c}%
\mathbf{\varphi}_{n-1}\\
\mathbf{\varphi}_{n-2}%
\end{array}
\right)  , \tag{7}\label{e8}%
\end{equation}
where $t$=$\frac{b+\sqrt{b^{2}+4(2+d)}}{2}$. More generally, we take
$\mathbf{\varphi}_{0}$ and $\mathbf{\varphi}_{L_{y}}$\ as the wave functions
at two open edges. Then we get a reduced transfer matrix linking the two edges
as follows:%

\begin{equation}
\left(
\begin{array}
[c]{c}%
\mathbf{\varphi}_{L_{y}+1}\\
\mathbf{\varphi}_{L_{y}}%
\end{array}
\right)  =M(\epsilon)\left(
\begin{array}
[c]{c}%
\mathbf{\varphi}_{1}\\
\mathbf{\varphi}_{0}%
\end{array}
\right)  , \tag{8}\label{e10}%
\end{equation}
where%

\begin{equation}
M(\epsilon)=\left[  \widetilde{M}(\epsilon)\right]  ^{L_{y}}=\left(
\begin{array}
[c]{cc}%
M_{11}(\epsilon) & M_{12}(\epsilon)\\
M_{21}(\epsilon) & M_{22}(\epsilon)
\end{array}
\right)  . \tag{9}\label{e11}%
\end{equation}
All kind of solutions of Eq. (\ref{e10}) are obtained by different choices of
$\mathbf{\varphi}_{0}$\ and $\mathbf{\varphi}_{1}$.

Similarly, one can obtain the eigenvalue equations of a graphene ribbon with
armchair edges,
\begin{align}
(\epsilon-M)\mathbf{\psi}_{nA}  &  =t_{1}e^{-ika}\mathbf{\psi}_{nB}%
+t_{1}e^{i\frac{ka}{2}}\left[  \mathbf{\psi}_{(n+1)B}+\mathbf{\psi}%
_{(n-1)B}\right] \nonumber\\
&  +t_{2}\left[  e^{-i\phi}\mathbf{\psi}_{(n+2)A}+e^{i\phi}\mathbf{\psi
}_{(n-2)A}\right] \nonumber\\
&  +\frac{t_{2}}{t_{1}}g_{0}\left[  e^{i\phi}\mathbf{\psi}_{(n+1)A}+e^{-i\phi
}\mathbf{\psi}_{(n-1)A}\right]  ,\nonumber\\
(\epsilon+M)\mathbf{\psi}_{nB}  &  =t_{1}e^{ika}\mathbf{\psi}_{nA}%
+t_{1}e^{-i\frac{ka}{2}}\left[  \mathbf{\psi}_{(n+1)A}+\mathbf{\psi}%
_{(n-1)A}\right] \nonumber\\
&  +t_{2}\left[  e^{i\phi}\mathbf{\psi}_{(n+2)B}+e^{-i\phi}\mathbf{\psi
}_{(n-2)B}\right] \nonumber\\
&  +\frac{t_{2}}{t_{1}}g_{0}\left[  e^{-i\phi}\mathbf{\psi}_{(n+1)B}+e^{i\phi
}\mathbf{\psi}_{(n-1)B}\right]  . \tag{10}\label{Armchair}%
\end{align}
However, because the derivation of the Harper equation for a graphene ribbon
with armchair edges is too sophisticated, here we do not write out the
transfer-matrix expression of this Harper equation. Moreover, because the main
results and the discussions on the graphene ribbons with zigzag and armchair
edges are similar, in the following we focus our attention to the graphene
with zigzag edges. The general open boundary condition is%

\begin{equation}
\mathbf{\varphi}_{L_{y}}=\mathbf{\varphi}_{0}=0. \tag{11}\label{e12}%
\end{equation}
With Eqs. (\ref{e10}) and (\ref{e11}), one can easily get that the solutions satisfy%

\begin{equation}
\ M_{21}(\epsilon)=0 \tag{12}\label{e13}%
\end{equation}
and%

\begin{equation}
\mathbf{\varphi}_{L_{y}-1}=-M_{11}\left(  \epsilon\right)  \mathbf{\varphi
}_{1}. \tag{13}\label{e14}%
\end{equation}
If we use a usual normalized wave function, the state is localized at the
edges as%
\begin{equation}
\left\{
\begin{array}
[c]{l}%
\left\vert M_{11}(\epsilon)\right\vert \ll1\text{ localized at }%
y\approx1\ \text{(down edge),}\\
\left\vert M_{11}(\epsilon)\right\vert \gg1\text{ localized at }y\approx
L_{y}-1\ \text{(up edge).}%
\end{array}
\right.  \tag{14}\label{e15}%
\end{equation}

Because the analytical derivation of the energy spectrum in the presence of
edges is very difficult, we now start a numerical calculation from Eqs.
(\ref{e3}) and (\ref{Armchair}). Varying all the controllable parameters,
which are the relative site energy $M/t_{1}$, the next nearest neighbor
hopping $t_{2}/t_{1}$, and the complex phase $\phi$, we can get three
different cases happening in the energy spectrum of the graphene ribbons. We
draw in Figs. 2(a)-(c) [Figs. 3(a)-(c)] the energy spectrum of graphene
ribbons with zigzag (armchair) edges as a function of $k_{x}$ for these three
different cases, i.e., the case $M/t_{2}<3\sqrt{3}\sin\phi$ (case I), the case
$M/t_{2}<-3\sqrt{3}\sin\phi$ (case II), and the case $M/t_{2}>3\sqrt{3}%
|\sin\phi|$ (case III), respectively. The number of sites A (B) in $y$
direction is chosen to be $L_{y}$=$40$. Clearly, from Figs. 2 and 3 one can
see that there are two dispersed energy bands (the shaded areas) with two edge
states (the colored lines) lying in the energy gap. It is our task to show
that the geometric nature of the edge states in these three kinds of parameter
regions are totally different, which can be described by the winding numbers
of the edge states on a complex energy RS within the topological edge theory
\cite{Hatsugai}. \begin{figure}[ptb]
\begin{center}
\includegraphics[width=1.0\linewidth]{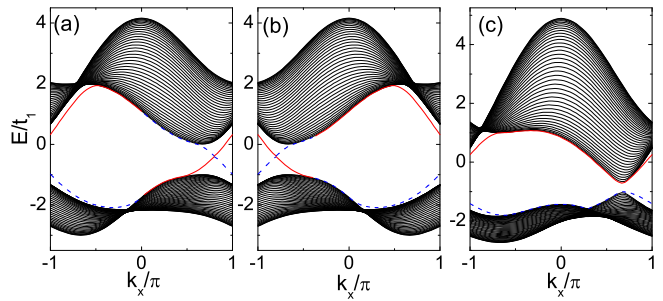}
\end{center}
\caption{(Color online) Energy spectrum of the graphene ribbon with zigzag
edges under different complex phases parameters: (a) $\phi$=$\pi/3$, (b)
$\phi$=$-\pi/3$, (c) $\phi$=$\pi/6$. The other parameters are set as $M/t_{1}%
$=$1$ and $t_{2}/t_{1}$=$1/3$ in all three figures. The shaded areas are the
energy bands and the colored lines are the spectrum of the edge states. The
red (solid) and blue (dashed) lines mean that the edge states are localized
near the down and up edges, respectively. }%
\end{figure}\begin{figure}[ptbptb]
\begin{center}
\includegraphics[width=1.0\linewidth]{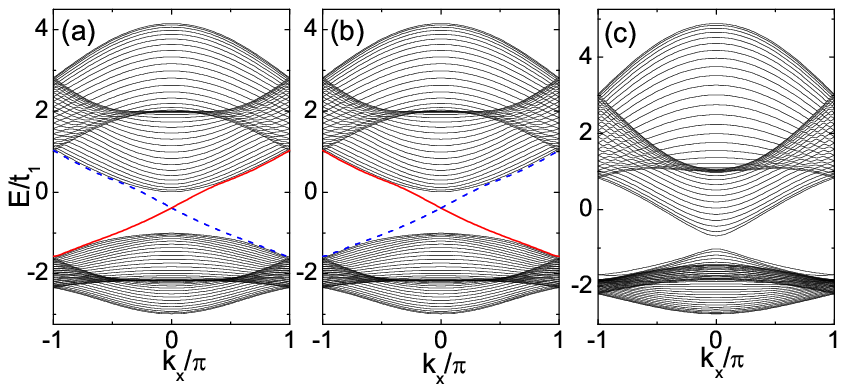}
\end{center}
\caption{(Color online) Energy spectrum of the graphene ribbon with armchair
edges under different complex phases parameters: (a) $\phi$=$\pi/3$, (b)
$\phi$=$-\pi/3$, (c) $\phi$=$\pi/6$. The other parameters are set as $M/t_{1}%
$=$1$ and $t_{2}/t_{1}$=$1/3$ in all three figures. The shaded areas are the
energy bands and the colored lines are the spectrum of the edge states. The
red (solid) and blue (dashed) lines mean that the edge states are localized
near the down and up edges, respectively. }%
\end{figure}

To show this, first, we ignore the open boundary condition and consider the
bulk Bloch function at sites with $y$-coordinate of $(L_{y}\mathtt{-}1)$. For
Bloch function, $\mathbf{\varphi}_{0}^{(b)}$\ and $\mathbf{\varphi}_{1}^{(b)}%
$\ compose an eigenvector of $M$\ with the eigenvalue $\rho$,%

\begin{equation}
M(\epsilon)\left(
\begin{array}
[c]{c}%
\mathbf{\varphi}_{1}^{(b)}\\
\mathbf{\varphi}_{0}^{(b)}%
\end{array}
\right)  =\rho\left(  \epsilon\right)  \left(
\begin{array}
[c]{c}%
\mathbf{\varphi}_{1}^{(b)}\\
\mathbf{\varphi}_{0}^{(b)}%
\end{array}
\right)  . \tag{15}\label{e16}%
\end{equation}
In order to discuss the wave function of the edge state, we extend the energy
to a complex energy. In the following, we use a complex variable $z$\ instead
of real energy$\ \epsilon$. From Eq. (\ref{e16}) we get%

\begin{equation}
\rho(z)=\frac{1}{2}\left[  \Delta(z)-\sqrt{\Delta^{2}(z)-4}\right]
\tag{16}\label{e17}%
\end{equation}
and%

\begin{equation}
\mathbf{\varphi}_{L_{y}-1}=-\frac{M_{11}\left(  z\right)  +M_{22}(z)-\omega
}{-M_{11}\left(  z\right)  +M_{22}(z)+\omega}M_{21}(z)\mathbf{\varphi}_{1},
\tag{17}\label{e18}%
\end{equation}
where $\Delta(z)$=Tr$\left[  M(z)\right]  $ and $\omega$=$\sqrt{\Delta
^{2}(z)-4}$. Clearly,%

\begin{equation}
\det M(\epsilon)=1 \tag{18}\label{e19}%
\end{equation}
since $\det\widetilde{M}(\epsilon)$=$1$. From Eq. (\ref{e18}) one can find
that the analytic structure of the wave function is determined by the
algebraic function $\omega$=$\sqrt{\Delta^{2}(z)-4}$. The RS of $\omega
$=$\sqrt{\Delta^{2}(z)-4}$\ on the complex energy plane can be built by the
conglutination between different analytic brunches. Here, the close complex
energy plane can be obtained from the open complex\ energy plane through
spheral pole mapping [see Fig. 4(a)]. Now let us discuss the analytic
structure of $\omega=\sqrt{\Delta^{2}(z)-4}$ on the open complex energy plane.
If the system has $q$\ energy bands, i.e.,%

\begin{equation}
\epsilon\in\left[  \lambda_{1},\lambda_{2}\right]  ,...,\left[  \lambda
_{2j-1},\lambda_{2j}\right]  ,...,\left[  \lambda_{2q-1},\lambda_{2q}\right]
, \tag{19}\label{e20}%
\end{equation}
where $\lambda_{j}$\ denote energies of the band edges and $\lambda
_{i}<\lambda_{j}$, $i<j$, then $\omega$\ can be factorized as%

\begin{equation}
\omega=\sqrt{\Delta^{2}(z)-4}=\sqrt{%
{\displaystyle\prod\limits_{j=1}^{2q}}
\left(  z-\lambda_{j}\right)  }.\tag{20}\label{e21000}%
\end{equation}
In the present Haldane model, there are two energy bands, so $q$=$2$. The two
single-value analytic branches are defined on the same complex energy plane
with $q$\ secants. The difference between the two branches are specified in
the following paragraph.

\begin{figure}[ptb]
\begin{center}
\includegraphics[width=1.0\linewidth]{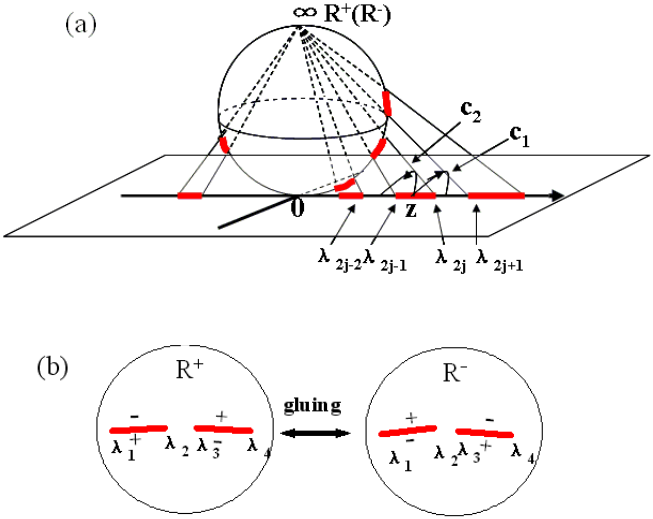}
\end{center}
\caption{(Color online) (a) The open complex energy plane are mapped to the
close complex energy plane through spheral pole projection. (b) Two sheets
with two cuts which correspond to the energy bands of the graphene
nanoribbons. The RS of the Bloch function is obtained by gluing the two
spheres along the arrows near the cuts. }%
\end{figure}

For an up- or down-edge-state energy $\mu_{j}$ in the gap $\left[
\lambda_{2j},\lambda_{2j+1}\right]  $, In order to ensure $\omega(\mu_{j}%
)\geq0$, we divide the two single-value analytic branches in terms of the
parity (evenness or oddness) of $j$. Let us consider the case that the energy
$z$ lies in the band $\left[  \lambda_{2j-1},\lambda_{2j}\right]  $ [see Fig.
4(a)]. The left side of this energy band is the\ $(j-1)$th\ gap, while the
right side is the $j$th gap. When the energy $z$\ moves from the $j$th band to
the $(j-1)$th gap (the $j$th gap) along an arbitrary path $c_{1}$ ($c_{2}$),
only the singularities $\lambda_{2j-1}$ and $\lambda_{2j}$\ have contributions
to the variance of the principal value of the argument of $\omega$. On the up
bank of the secant, we distinguish two branches $R^{+}$\ and $R^{-}$ as the
following: For even values of $j$, if we set $\arg(z-\lambda_{2j-1})$=$0$ and
$\arg(z-\lambda_{2j})$=$\pi$, which corresponds to $\omega(\mu_{j-1})>0$
($\omega(\mu_{j})<0$) when $z$ moves along $c_{1}$ ($c_{2}$), then the branch
$R^{+}$ is defined as a complex plane with $q$ secants. Whereas, if we set
$\arg(z-\lambda_{2j-1})$=$2\pi$ and $\arg(z-\lambda_{2j})$=$\pi$, which
corresponds to $\omega(\mu_{j-1})<0$ ($\omega(\mu_{j})>0$) when $z$ moves
along $c_{1}$ ($c_{2}$), then the branch $R^{-}$ is defined as a complex plane
with $q$ secants. The definitions of $R^{+}$ and $R^{-}$ for odd values of $j$
are reverse to those for even values of $j$. So, if $z$\ lies in the $j$th gap
from below on the real axis,%

\[
\alpha\left(  -1\right)  ^{j}\omega\geq0,\ z\text{:}\ \text{real on }%
R^{\alpha}\ (\alpha=+,-),
\]
and at energies $\mu_{j}$ ($\mathtt{\in}R^{\alpha}$, $\alpha$=$+,-$) of the
edge states we have%

\begin{equation}
\omega(\mu_{j})=\alpha\left(  -1\right)  ^{j}\left\vert M_{11}\left(  \mu
_{j}\right)  -M_{22}\left(  \mu_{j}\right)  \right\vert , \tag{21}\label{e25}%
\end{equation}
In addition, one can easily obtain%

\begin{equation}
\Delta(\epsilon)\left\{
\begin{array}
[c]{c}%
\leq-\text{2 for }j\text{ odd}\\
\geq\text{2 for }j\text{ even}%
\end{array}
\right.  , \tag{22}\label{e26}%
\end{equation}
where the energy $\epsilon$\ (on $R^{\alpha}$)\ is in the $j$th\ gap.

When the branches $R^{+}$ and $R^{-}$\ on the open complex energy plane are
mapped to the close complex energy plane through spheral-pole-projection, one
can get two single-value analytic spherical surfaces. The RS is obtained by
gluing the two spherical surfaces at these branch cuts with making sure that
the $\pm$ banks face the $\mp$ banks of other sphere [see Figs. 4(b)]. Note
that there are two real axes after gluing. In the present model the genus of
the RS is $g$=$1$, which is the number of energy gaps. The wave function is
defined on the $g$=$1$ RS $\Sigma_{g=1}(k_{x})$. The branch of the Bloch
function is specified as $\omega\mathtt{>}0$, which we have discussed above.
With Eqs. (\ref{e26}), (\ref{e25}) and (\ref{e18}), and using the fact that
$\mathbf{\varphi}_{L_{y}-1}(\mu_{j})\mathtt{=}0$\ for $\mu_{j}\mathtt{\in
}R^{\alpha}$ and $\mathbf{\varphi}_{L_{y}-1}(\mu_{j})\mathtt{\neq}0$\ for
$\mu_{j}\mathtt{\in}R^{-\alpha}$, one can obtain that when the zero point is
on the upper sheet of RS ($R^{+}$), the edge state is localized at the down
edge; when the zero point is on the lower sheet of RS ($R^{-}$), the edge
state is localized at the up edge. \begin{figure}[ptb]
\begin{center}
\includegraphics[width=0.8\linewidth]{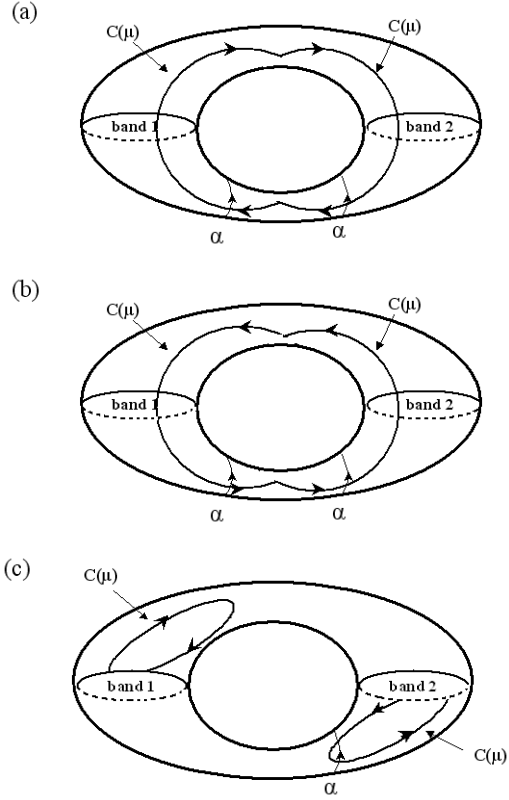}
\end{center}
\caption{Riemann surfaces of the Bloch functions for different winding
numbers: (a) $I$=$1$, (b) $I$=$-1$, and (c) $I$=$0$.}%
\end{figure}

Figure 5(a)-(c) schematically show the RS's for the present Haldane model with
the system parameters belonging to case-I ($M/t_{2}<3\sqrt{3}\sin\phi$),
case-II ($M/t_{2}<-3\sqrt{3}\sin\phi$), and case-III ($M/t_{2}>3\sqrt{3}%
|\sin\phi|$), respectively. On each RS $\Sigma_{g=1}(k_{x})$ the energy gap
corresponds to the loop around the hole of the $\Sigma_{g=1}(k_{x})$ and the
energy bands correspond to the closed paths vertical to the energy gap loop on
$\Sigma_{g=1}(k_{x})$. The Bloch function is defined on this surface. For the
fixed $k_{x}$ and $\phi$, there is always $g$=$1$\ zero point at the
down-edge-state energy $\mu_{j}^{(\text{down})}$. Since there are two real
axes on the $\Sigma_{g=1}(k_{x})$, correspondingly, there is $g$=$1$\ zero
point at the up-edge-state energy $\mu_{j}^{(\text{up})}$.

The above analysis is for the fixed $k_{x}$. Changing $k_{x}$ in one period,
we can consider a family of RS's $\Sigma_{g=1}(k_{x})$. $\Sigma_{g=1}(k_{x})$
can be modified by this change. However, all the RS's $\Sigma_{g=1}(k_{x})$
with different $k_{x}\ $are topologically equivalent if, as what happens in
the present model, there are stable energy gaps in the 2D energy spectrum. By
identifying the topologically equivalent RS's $\Sigma_{g=1}(k_{x})$, the
behavior of the track of $\mu_{j}(k_{x})$ (including the up-edge-state energy
$\mu_{j}^{(\text{up})}$ and down-edge-state energy $\mu_{j}^{(\text{down})}$)
depends on system parameters, as shown in Fig. 5. In Fig. 5(a), which
corresponds to case-I of $M/t_{2}<3\sqrt{3}\sin\phi$, one can observe that by
varying $k_{x}$, the down-edge-state energy $\mu_{j}^{(\text{down})}(k_{x})$
moves from the lower band (band 1 in Fig. 5) edge to the upper band (band 2 in
Fig. 5) edge, while the down-edge-state energy $\mu_{j}^{(\text{down})}%
(k_{x})$ moves from the upper band edge to the lower band edge. That is to
say, the two edge-state energy tracks in the same energy gap move around the
hole and form an oriented loop $C(\mu_{j})$. In case-II of $M/t_{2}%
\mathtt{<}-3\sqrt{3}\sin\phi$, as shown in Fig. 5(b),\ the two edge-state
energy tracks moving around the hole also form an oriented loop. However, the
orientation of the loop is right about with respect to that in case-I.
Finally, in case-III of $M/t_{2}\mathtt{>}3\sqrt{3}|\sin\phi|$, as shown in
Fig. 5(c), one can observe that the $\mu_{j}(k_{x})$\ moves along the hole and
turns back before arriving at the second energy band. In this case, the two
edge-state energy tracks in the same energy gap approximately form two
circularities.\begin{figure}[ptb]
\begin{center}
\includegraphics[width=0.8\linewidth]{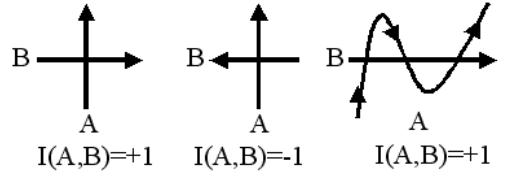}
\end{center}
\caption{Intersection number $I(A,B)$ of two curves A and B. Each intersection
point contributes by $+1$ or $-1$ according to the direction.}%
\end{figure}

It is known that on a general genus-$g$\ RS,\ all kinds of loops (the first
homotopy group) are generated by $2g$\ canonical loops (generators),
$\alpha_{i}$\ and $\beta_{i}$, $i$=$1,\ldots,g$. See Fig. 5 for $g\mathtt{=}%
1$. The intersection number of these curves (including directions)
\cite{Hatsugai} is given by (see Fig. 6)%

\begin{equation}
I\left(  \alpha_{i},\beta_{j}\right)  =\delta_{ij}. \tag{23}\label{e27}%
\end{equation}
Any curves on the RS are spanned homotopically by $\alpha_{i}$\ and $\beta
_{i}$. When the edge-state energy loop $\mu_{j}(k_{x})$\ moves $p$\ times
around the $j$th\ hole with some integer $p$, one has%

\begin{equation}
C(\mu_{j})\approx\beta_{j}^{p}, \tag{24}\label{e28}%
\end{equation}
which means%

\begin{equation}
I\left(  \alpha_{i},C(\mu_{j})\right)  =p\delta_{ij}. \tag{25}\label{e29}%
\end{equation}

When the Fermi energy $\epsilon_{F}$\ of the 2D system lies in the $j$th bulk
energy gap, the Hall conductance is given by the winding number of the edge
state \cite{Hatsugai}, which is given by the number of intersections
$I(\alpha_{j},C(\mu_{j}))$ ($\equiv I(C(\mu_{j}))$)\ between the canonical
loop $\alpha_{j}$\ on the RS and the trace of $\mu_{j}$. In the present
single-gap model, we obtain the Hall conductance provided by the edge states
as follows%

\begin{equation}
\sigma_{xy}^{\text{edge}}=-\frac{e^{2}}{h}I(C(\mu)). \tag{26}\label{e30}%
\end{equation}
From Figs. 5(a) and (b) it can be observed that $\mu$\ moves one time across
the hole ($p\mathtt{=}1$), which in terms of Eq. (\ref{e28}) means $C(\mu
)$\ $\approx\beta$. Considering simultaneously the winding direction (see Fig.
6), one can obtain that in Fig. 5(a) $I(C(\mu))$=$-1$,\ while in Fig. 5(b)
$I(C(\mu))$=$1$. In Fig. 5(c), because $p$=$0$, $I\left(  \alpha_{i},C(\mu
_{i})\right)  $=$0$ and the Hall conductivity is zero. Therefore, we get%

\begin{equation}
\sigma_{xy}^{\text{edge}}=\left\{
\begin{array}
[c]{cc}%
\frac{e^{2}}{h}, & M/t_{2}<3\sqrt{3}\sin\phi\\
0, & M/t_{2}>3\sqrt{3}|\sin\phi|\\
-\frac{e^{2}}{h}, & M/t_{2}<-3\sqrt{3}\sin\phi
\end{array}
\right.  . \tag{28}\label{e32}%
\end{equation}

Finally, let us compare this result for the graphene ribbons with
zigzag/armchair edges with the bulk graphene, in which the topological
invariant is the Chern number. In the bulk Haldane model, when the Fermi
energy $\epsilon_{F}$\ lies in the energy gap, the Hall conductance is
quantized as $\sigma_{xy}^{\text{bulk}}=\frac{e^{2}}{h}C_{1}$, where $C_{1}$
is the Chern number of the lower energy band. It turns out that%

\begin{equation}
C_{1}=\left\{
\begin{array}
[c]{cc}%
1, & M/t_{2}<3\sqrt{3}\sin\phi\\
0, & M/t_{2}>3\sqrt{3}|\sin\phi|\\
-1, & M/t_{2}<-3\sqrt{3}\sin\phi
\end{array}
\right.  . \tag{29}\label{e33}%
\end{equation}
From Eqs. (\ref{e32}) and (\ref{e33}) one can obtain $\sigma_{xy}%
^{\text{edge}}=\sigma_{xy}^{\text{bulk}}$, which is in accord with the
established recognition \cite{Halperin,Hatsugai} on the Hall conductance in
the systems with and without edges.

In summary, we have investigated the topological property of the edge states
in the Haldane model. The Harper equations for solving and analyzing the edge
states have been derived. It has been found that there are two edge states
lying in the bulk energy gap. These two edge states move with varying $k_{x}$
around the hole in the RS and form an orientated energy loop. With the winding
number of the edge states, we have obtained that the edge-state Hall
conductance is $\sigma_{xy}^{\text{edge}}$=$\pm\frac{e^{2}}{h}$ or $0$ under
different cases, which agrees with that based on the topological bulk theory.

\end{document}